\documentclass[manuscript]{acmart}
\renewcommand\footnotetextcopyrightpermission[1]{}
\usepackage{xspace}
\usepackage{mdframed}
\usepackage{color, colortbl}
\usepackage{array}
\usepackage{scalefnt}
\usepackage{comment}
\usepackage{xcolor}
\usepackage{placeins}

\AtBeginDocument{%
  \providecommand\BibTeX{{%
    \normalfont B\kern-0.5em{\scshape i\kern-0.25em b}\kern-0.8em\TeX}}}

\setcopyright{acmlicensed}
\acmJournal{PACMHCI}
\acmYear{2022} \acmVolume{6} \acmNumber{CSCW2} \acmArticle{330} \acmMonth{11} \acmPrice{15.00}\acmDOI{10.1145/3555221}

\begin{document}

\title{The Future of Hackathon Research and Practice}

\author{Jeanette Falk}
\email{jeanette.falk-olesen@plus.ac.at}
\affiliation{%
  \institution{University of Salzburg}
  \city{Salzburg}
  \country{Austria}
}

\author{Alexander Nolte}
\email{alexander.nolte@ut.ee}
\affiliation{%
  \institution{University of Tartu}
  \city{Tartu}
  \country{Estonia}
}
\affiliation{%
  \institution{Carnegie Mellon University}
  \city{Pittsburgh}
  \state{PA}
  \country{USA}
}

\author{Daniela Huppenkothen}
\email{d.huppenkothen@sron.nl}
\affiliation{%
  \institution{SRON Netherlands Institute for Space Research}
  \city{Leiden}
  \country{The Netherlands}
}

\author{Marion Weinzierl}
\email{marion.weinzierl@durham.ac.uk}
\affiliation{%
  \institution{Durham University}
  \city{Durham}
  \country{United Kingdom}
}

\author{Kiev Gama}
\email{kiev@cin.ufpe.br}
\affiliation{%
  \institution{Federal University of Pernambuco}
  \city{Recife}
  \state{PE}
  \country{Brazil}
}

\author{Daniel Spikol}
\email{ds@di.ku.dk}
\affiliation{%
  \institution{University of Copenhagen}
  \city{Copenhagen}
  \country{Denmark}
}

\author{Erik Tollerud}
\email{etollerud@stsci.edu}
\affiliation{%
  \institution{Space Telescope Science Institute}
  \city{Baltimore}
  \state{MD}
  \country{USA}
}

\author{Neil Chue Hong}
\email{n.chuehong@software.ac.uk}
\affiliation{%
  \institution{University of Edinburgh}
  \city{Edinburgh}
  \country{Scotland}
}

\author{Ines Knäpper}
\email{ines.knaepper@weforum.org}
\affiliation{%
  \institution{World Economic Forum}
  \city{Geneva}
  \country{Switzerland}
}

\author{Linda Bailey Hayden}
\email{ines.knaepper@weforum.org}
\affiliation{%
  \institution{Elizabeth City State University}
  \city{Elizabeth City}
  \state{NC}
  \country{USA}
}

\renewcommand{\shorttitle}{The Future of Hackathon Research and Practice}
\renewcommand{\shortauthors}{Falk and Nolte et al.}

\begin{abstract}
Hackathons are time-bounded collaborative events which have become a global phenomenon adopted by both researchers and practitioners in a plethora of contexts. Hackathon events are generally used to accelerate the development of, for example, scientific results and collaborations, communities, and innovative prototypes addressing urgent challenges. As hackathons have been adopted into many different contexts, the events have also been adapted in numerous ways corresponding to the unique needs and situations of organizers, participants and other stakeholders. While these interdisciplinary adaptions, in general affords many advantages --- such as tailoring the format to specific needs --- they also entail certain challenges, specifically: 1) limited exchange of best practices, 2) limited exchange of research findings, and 3) larger overarching questions that require interdisciplinary collaboration are not discovered and remain unaddressed. We call for interdisciplinary collaborations to address these challenges. As a first initiative towards this, we performed an interdisciplinary collaborative analysis in the context of a workshop at the Lorentz Center, Leiden in December 2021. In this paper, we present the results of this analysis in terms of six important areas which we envision to contribute to maturing hackathon research and practice: 1) hackathons for different purposes, 2) socio-technical event design, 3) scaling up, 4) making hackathons equitable, 5) studying hackathons, and 6) hackathon goals and how to reach them. We present these areas in terms of the state of the art and research proposals and conclude the paper by suggesting next steps needed for advancing hackathon research and practice.
\end{abstract}


\maketitle

\section{Introduction}
Time-bounded collaborative events have become a global phenomenon. 
They are referred to as hackathons, hack weeks, hack-days, data dives, codefests, sprints and others.
They started as niche competitive events in the early 2000s during which mostly junior developers formed small ad-hoc teams to work on a software project for pizza and sometimes the prospect of a future job. 
Since then they have moved into a plethora of contexts including science~\cite{huppenkothen2018hack}, industry~\cite{lewis2015ux},  entrepreneurship~\cite{cobham2017appfest}, government institutions~\cite{briscoe2014digital}, non-profit organizations~\cite{trinaistic2020hackathons}, education~\cite{wang2018extended}, civic engagement~\cite{taylor2018everybody}, and libraries~\cite{longmeier2021hackathons}, and involving fields such as design, computer science, arts, health and marketing, to mention a few. 
The largest hackathon database\footnote{\href{https://devpost.com/hackathons}{https://devpost.com/hackathons}} lists almost 1200 events that took place in 2021 alone with most of them running under the umbrella of Major League Hacking (MLH)~\footnote{\href{https://mlh.io/}{https://mlh.io/}}. 
Since MLH only registers some hackathons and mainly focuses on North America and Europe it can be expected though that the actual number of annual hackathon events is much larger.

Most hackathon events share commonalities like being collaborative, time-bounded and motivated by an overarching theme, which participants aim to address.
The adoption of hackathons in different domains has consequently led to the formats evolving in different directions to suit specific needs or foster specific goals. 
Therefore, apart from the above-mentioned commonalities, there is an abundance of different approaches to almost any other aspect of the format including but not limited to: the length and size of events, how participants are recruited, how their collaboration is structured, which support they receive, which tools and materials participants work with and have access to, how these tools and materials are introduced to participants and so on. 

We observe a similar variety and specialization in the context of scientific literature covering hackathons. 
Research focusing on these events has developed in various disciplines and domains including computer science, high performance computing, astrophysics, and others. 
Research on such events has also started somewhat late. 
While the term hackathon first emerged in the early 2000s, the first research papers covering the topic did not emerge until 5 to 10 years later~\cite{falk202010}.

While we acknowledge the advantages of interdisciplinarity when it comes to organizing and studying hackathon events, we also observe that both research and practice around hackathons is quite \textit{siloed}. 
Different communities organize and study events with little exchange between them which poses significant risks, such as repeating poor or, in worst cases, even harmful practices, leading to sub-optimal experiences and outcomes for not only researchers but also for participants, organizers and connected stakeholders.
These poor practices include repeating and amplifying not only superficial but even technosolutionist approaches to complex social and societal issues~\cite{falk2021hackathons}, or inadvertently contribute to bad experiences or limited access of participation for minoritized people, wishing to participate~\cite{paganini2020}.

Advancing theory and practice for interdisciplinary researchers who study the same phenomena of hackathons is challenging if awareness of relevant and prior theory is lacking because of siloed research and practice. 
This can create problems that ``diminish the effectiveness of the research products''~\cite{edmondson2007methodological}, if there is not an appropriate alignment between research question, prior work, research design and contribution to literature ~\cite{edmondson2007methodological}.
In addition to our practical experiences, our observations of poor methodological fit in terms of siloed research is informed by our experience from reviewing research, similar to ~\cite{edmondson2007methodological}.
In section \textit{Distribution of research on hackathons}, we confirm this observation quantitatively using bibliometric tools.

We argue that \textit{siloing} and the resulting poor methodological fit of hackathon research and practice inhibits progress in both areas. 
In particular it results in the following challenges:
\begin{enumerate}
    \item \textbf{Limited exchange of best practices} which leads to organizers having to rediscover the same things, repeating mistakes and the format overall drifting into different directions without critical reflection.
    \item \textbf{Limited exchange of research findings} which leads to repeating studies that discover the same or similar things resulting in a stagnation of research progress.
    \item \textbf{The biggest research and practice challenges} that require interdisciplinary collaboration are not discovered and \textbf{remain unaddressed}.
\end{enumerate}

To address these challenges, we organized a workshop at the Lorentz Center\footnote{\href{https://bit.ly/3gtv4Gl}{https://bit.ly/3gtv4Gl}}. 
We brought together hackathon researchers and practitioners from various disciplines including software engineering, high-performance computing, information systems, astronomy, geology, physics and organizational sciences. 

This paper presents the result of a multidisciplinary analysis of hackathon research and practice during that workshop. 
First, we start the article by outlining recent work that provides overviews and reviews of hackathon research and practice.
We use these as a point of departure for continuing the improvement of understanding hackathons and as motivation for our focus on an interdisciplinary perspective on unified problem formulations, sharing of methods and creation of new research questions~\cite{miller2008epistemological,eigenbrode2007employing}. 
Second, to support our observation of how hackathon research seems to be siloed, we provide a network analysis of the distribution of hackathon research.
Third, since research on hackathon is tightly connected to the practical sphere of hackathon organization, we discuss some logistics and facilitation of hackathon organization based on our experience of hackathon practice.
This section is particularly suited to readers who wish to organize hackathons. 
Fourth, we envision directions for future hackathon research and practice structured into six areas:
(1) Hackathons for different purposes, 
(2) Socio-technical event design, 
(3) Scaling up, 
(4) Making hackathons equitable, 
(5) Studying hackathons, and
(6) Hackathon goals and how to reach them.

The purpose of this paper is, first, to support researchers in the pursuit of future research endeavors. 
Second, we aim to help practitioners to identify areas in the organization of hackathons to explore and further develop the format, and to improve participation (diversity and inclusion of participants, mentoring, support, etc.).

\section{Background}
\label{sec:background}
Some notable examples on publications have recently synthesized insights on hackathon research and practice, including literature reviews and case studies:
For example, Flus and Horst~\cite{flus2021design} conduct a literature review focusing on characterizing the design activity at hackathons and discussing future design research on hackathons.
Falk Olesen and Halskov's~\cite{falk202010} literature review uses the Association for Computing Machinery (ACM) Digital Library (DL) to study the relationship between research and hackathons and provide an overview of challenges and opportunities within this relationship.
Kollwitz and Dinter~\cite{kollwitz2019hack} systematically review 189 research publications with a focus on the Information System research domain at ``the crossroads of digital innovation and OI (open innovation)''~\cite{kollwitz2019hack} and on this basis develop a taxonomy of hackathon dimensions.

Overviews of hackathon practice include Rys'~\cite{rys2021invention} evaluation of 14 hackathons as an invention development method compared to brainstorming, where they explore how hackathons may mitigate some of the drawbacks of brainstorming. 
Pe-Than et al.~\cite{pe2019designing} review ten hackathons and research literature to discuss design choices which hackathon organizers should consider. 
Nolte et al.~\cite{nolte2020organize} extend this work and develop an online kit to support hackathon organization.

These papers are notable contributions to the development of an ontology of hackathons and to uncovering best practices for organization.
While they are important in developing an understanding of hackathons and how they can be organized in different contexts, the studies have generally been conducted \textit{within} disciplines. 
In this paper, we offer an interdisciplinary perspective on the phenomenon of hackathons as well as move towards outlining paths for formulating unified problems across disciplines, sharing of methods, and developing new research questions~\cite{miller2008epistemological,eigenbrode2007employing}.

\subsection{Defining Hackathons}
To place hackathons into an interdisciplinary context first requires a framework for what constitutes a hackathon.
As we established earlier in this paper and explain in more depth in the next section, hackathons are conducted in many different contexts and fields and with different purposes.
This suggest that finding a unified, interdisciplinary definition may be a complex problem.

Our aim is to propose a framing that is inclusive rather than exclusive. 
It should include any event that can conceivably be perceived as a hackathon. 
At the same time it should not be so broad that it could include any event where people come together such as concerts, conventions, workshops or similar. 
The traits we describe in the definition below are common traits for many if not most relevant hackathon events, but might be different for specific hackathons that are designed for a specific purpose.

We \textbf{define hackathons} as time-bounded participant-driven events that are organized to foster specific goals or objectives. 
The scaffolding of each event is planned by a team of organizers to support its goals or objectives. 
People that participate in an event often (but not necessarily) have different backgrounds and bring different expertise. 
Their primary motivation to join an event is to work on a shared team project that interests them, although there might be additional incentives such as prizes and networking opportunities. 
During the event, teams attempt to create an artifact (e.g. software or hardware prototypes, slides, video, document) that can be shared with other participants. 
It is also acceptable, and sometimes even desirable, if they do not manage to create anything. 
Participants are encouraged to be bold and work on things outside of their area of expertise.

In the next subsection, we outline some fields of study which hackathon research and practice connect to.

\subsection{Hackathons in context}
Hackathons at their core are collaborative events. 
Collaboration in these events takes place mainly in small teams that work independently~\cite{trainer2016hackathon}. 
There is much work in different domains that focuses on how (small) teams communicate and collaborate, including psychology, education, organizational sciences, volunteer engagement, entrepreneurship and others. 
This work is deeply relevant to the study of hackathons since we expect teams to face similar challenges related to communication, organization, leadership, equity and others. 
However, two major defining factors in hackathons are their time-bounded nature, and the feature that team members might meet each other at the event for the first time. 
As a result, teams have to establish how they collaborate in a short period of time --- related to what Edmondson has coined ``teaming''~\cite{edmondson2012teaming} --- which has been described as a key characteristic of hackathon participation~\cite{falk2022supporting}.

Collaboration in teams during a hackathon revolves around a project chosen by teams themselves~\cite{pe2019understanding}. 
These projects often focus on the creation of a (technical) artifact, like a web site, mobile application, robot or a piece of software. 
Our understanding of how teams engage with this task can be informed by work related to project management, agile software engineering, design and others. 
Differences between those works and hackathons relate e.g. to the way that teams approach their projects. 
In hackathons team members often choose tasks within their projects that they are interested in rather than tasks that correspond to their individual skill-set~\cite{nolte2018you} -- often they might even choose a task they have no prior knowledge in, as they use the hackathon as a learning opportunity. 
Teams also often approach projects without any or limited prior planning and engage in a form of rapid prototyping.

While teams mainly work in an independent and self-directed manner, their collaboration still takes place in the context of a specific event. 
The planning of the overall event consequently influences teams in particular in the context of team formation and project ideation~\cite{olesen2018four}. 
Organizers often deploy means of facilitation to keep teams on track~\cite{taylor2018everybody}. 
Related works on other collaborative settings such as game jams, workshops, classrooms and teamwork may support our understanding of how teams engage in such environments. 
A key difference between hackathons and other events is a much looser scaffolding where fewer rules are enforced.

\section{Distribution of research on hackathons}
\label{Distribution}
We analyzed research outputs around hackathons to get an overview of which research disciplines are involved and how authors collaborate, to illustrate the siloed effect we discussed in the introduction. 
For this purpose we used bibliometric tools, which can be used to analyze the impact of a research area~\cite{jefferson2017comment}. 
We used the Lens Scholarly Analysis tool~\cite{RN310}, which uses global public resources like PubMed, MAG, and Crossref for science and innovation assessment. 
Using the search (title:(Hackathon*) OR abstract:(Hackathon*) OR keyword:(Hackathon*) OR field-of-study:(Hackathon*)) yielded 1,794 scholarly works and 651 scholarly works cited by other literary works\footnote{We performed the search on July 7, 2022. The full dataset is available here: \href{https://bit.ly/3nH6C7T}{https://bit.ly/3nH6C7T}}.
While this gives a good overall trend of rise of hackathon numbers, it doesn't measure e.g. papers studying adjacent collaborative events and may misrepresent some papers looking at hackathons by different names such as coding sprints.

\begin{figure*}
\centering
\includegraphics[width=\linewidth]{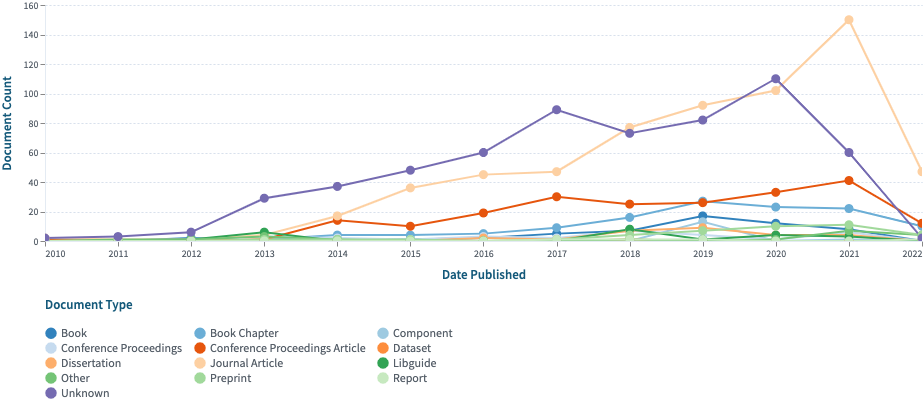}
\caption{Publications over time from 2010-2021}
\label{fig:journals}
\vspace{-10pt}
\end{figure*}

Our findings indicate that interest in hackathons has steadily risen over the past ten years. 
Fig.~\ref{fig:journals} particularly shows a continuing rise of journal articles and conference proceedings. 
It also shows that non-peer-reviewed articles have started a few years earlier compared to journal articles and conference proceedings and been the most common form of publication until 2020.
Since then, they have started to decline while other forms of publication continue to grow providing indication for the field maturing. 

\begin{figure}
\centering
\includegraphics[width=\linewidth]{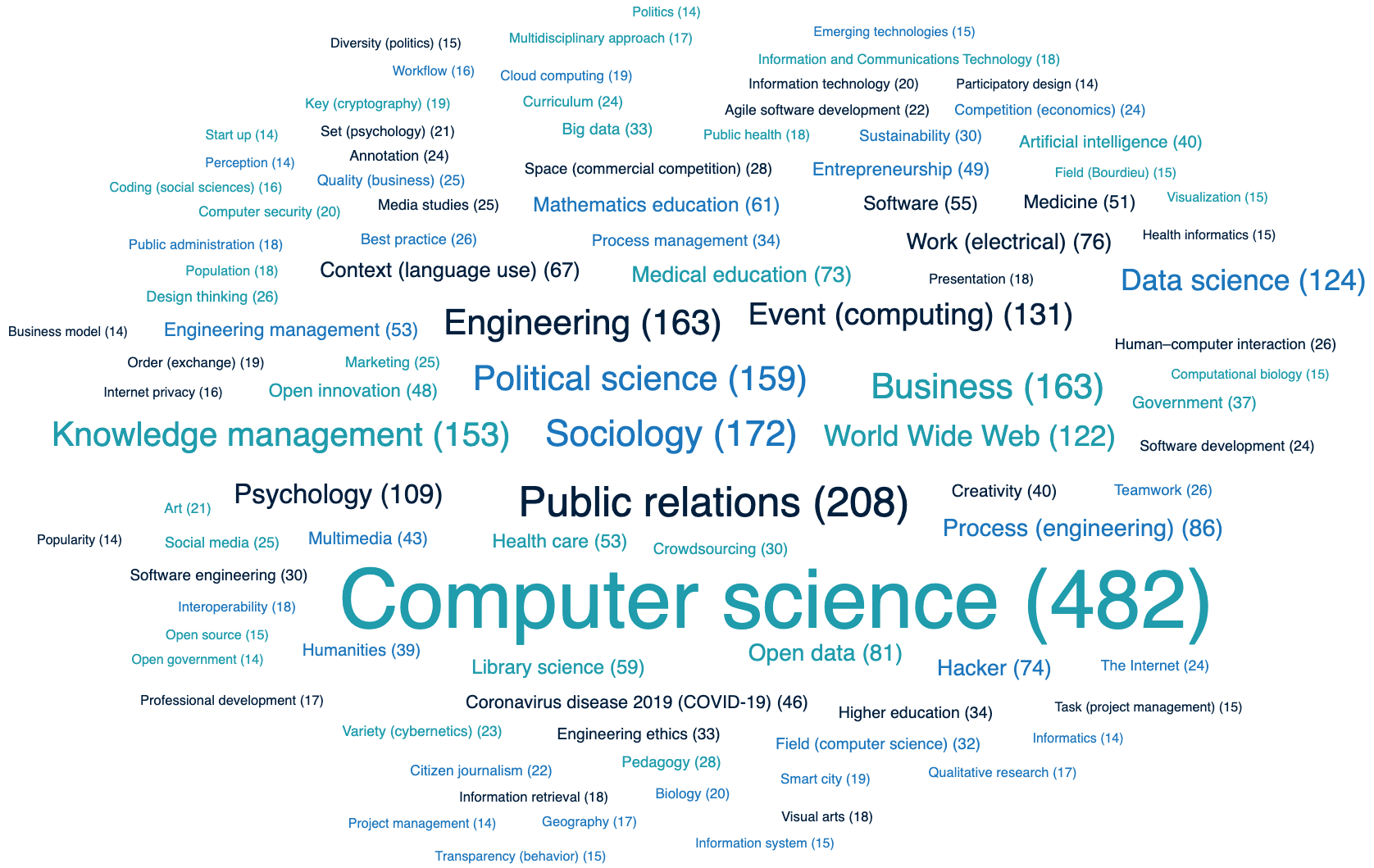}
\caption{Word-cloud of fields of study}
\label{fig:wordcloud}
\vspace{-10pt}
\end{figure}

To show the diversity of fields that organize and study hackathons we constructed a word cloud (Fig.~\ref{fig:wordcloud}) based on a service provided by Microsoft Academic that uses machine learning parsing of all accessible text in the record (title, abstract, and keywords). 
It illustrates that computer science is mentioned most (482 articles) followed by public relations (208). 
We also analyzed the citation metrics for the papers in our corpus and found that the most highly cited paper is in computer science (155 citations). 
The second and third most cited papers are a paper around informal learning (113 citations) and a paper that focuses on intense collocation and collaboration (89 citations). 
The next field of study is public relations and the three most cited papers in public relations are related to civic hackathons (170 citations) and entrepreneurial citizenship (162 citations). 
While in computer science, the most cited paper is from computational linguistics and mental health (162 citations) and the second paper is hackathons on informal learning (117 citations).

\begin{figure*}
\centering
\includegraphics[width=\linewidth]{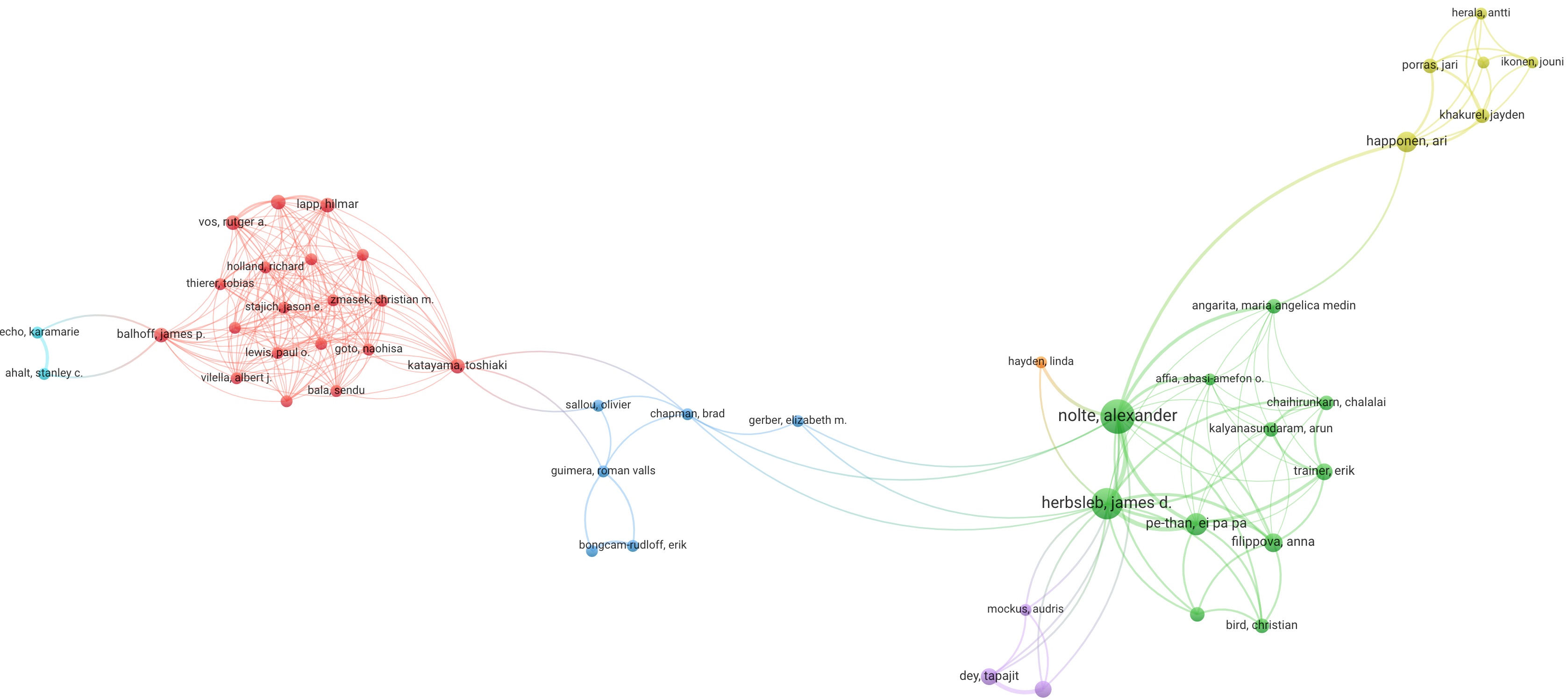}
\caption{Author map of contributors to hackathon research as defined in the text. This clearly shows how siloed this research area is, as the clusters are not well-connected.}
\label{fig:authors}
\vspace{-10pt}
\end{figure*}

Finally, we also conducted a co-authorship analysis. 
For this we used VOSview~\cite{van2010software} with a RIS export from the Lens Scholarly Analysis with a minimum number of documents of an author set to 2 with fractional counting (weighted), which set a threshold from the 2839 authors to 470 that meet the threshold. 
We visualized these 470 co-authors in Fig.~\ref{fig:authors}. 
This illustrates clearly how siloed the research in this area is -- while there are some connections between the clusters, the clustering is far more prominent than the inter-cluster connections.

\section{Hackathon organization}
\label{hackathonOrganization}
As described in the previous section, hackathons started as hands-on collaborative events, and it took a while for research to take in interest in them. 
Therefore, our foundation for the paper is not only informed by the emerging research on hackathons, but also to a great extent by the practice of hackathon organization and participation.
This section establish the foundations which we are starting from, by summarizing key aspects of logistics and facilitation of organizing hackathons. 
Most of these aspects are experience-based and anecdotal, rather than based on studies. 
The next section will touch on some of this, and point out where formal studies might be beneficial and lead to insights that can improve hackathon organization.

A key element of hackathons is getting people out of their ``day job'', so the event format must be different from what they are used to in their day-to-day work. 
This is often achieved by creating a relatively informal atmosphere, allowing for spontaneity and bringing in of unexpected ideas. 
Thus, the organizers and participants need to be prepared to constantly adjust their plans to what is happening at the event. 
The community-developed online resource called the ``Hackathon Planning Kit\footnote{\href{https://hackathon-planning-kit.org/}{https://hackathon-planning-kit.org/}}''~\cite{nolte2020organize} is useful to get started with organizing a hackathon and to make sure that no important aspect is forgotten. 

Hackathons are inherently experimental. 
The participants are allowed and encouraged to experiment and fail, rather than following a set step-by-step tutorial to fulfill a given task. 
In the same way, hackathon organizers should allow themselves to experiment and to try out new things.

\subsection{Logistics of Hackathon Settings}
Due to the agile nature of hackathons, flexible spaces are required which can be adjusted to the needs of participants. 
For in-person events, adjustable rooms, movable tables and chairs, lots of boards and screens, enough power plugs and extension cables, wifi, and plenty of space, based on the number of participants, are required. 
An example on flexible spaces is the UW Active Learning Classroom~\cite{UW}, while setups like traditional lecture theatres should be avoided.
For online hackathons, the flexibility needs to be reflected in the range of tools used, for example Zoom for plenary events, Zoom breakout rooms for the hackathon work, and a persistent chat tool with configurable channels, such as Slack or Discord, for asynchronous communication. 
A central notice board or shared documents with links and all necessary information to know where you have to be and where things are, who is in which group, etc., can be very helpful to avoid people getting lost in this multitude of material and spaces. 
In order to keep teams together and make sure people do not forget to look after themselves, breaks should be scheduled centrally, and at in-person events food should be provided in-house. 
This is also a chance to provide networking opportunities. 
Participants can also be brought together for tutorials or talks, teaching technologies and methods relevant to the hackathon, or providing insight into an overarching theme.

To make sure that, in spite or because of all the flexibility, the organizers do not get carried away and lose touch with the participants, hackathons are often evaluated in-depth.
For example, some hackathons have physical or virtual feed back walls during the entire event in addition to post-event surveys. 
While it is not possible to use this evaluation to do "everything right next time", it is an important tool for reflecting on what worked and what didn't, and to make more informed decisions for future events.
    
\subsection{Facilitation of Hackathon Participation}
Hackathons generally have less of a strict schedule, are spontaneous, and therefore also require an increased flexibility from organizers and the ability to adjust and adapt quickly. 
However, designing facilitation plans for use throughout the event lowers the mental load on organizers during the hackathon itself.
This is not to say that there is necessarily only a low upfront effort before the event: organizers who e.g. emphasize inclusivity in their hackathons may also spend considerable time and energy into planning appropriate physical and digital spaces as well as processes.  
For example, considering the latter, some hackathons implement active participant selection processes in order to create a cohort diverse in background and experience, which in turn requires the design and implementation of selection processes.
One online resource which may assist hackathon organizers in participant selection processes is \textit{entrofy}~\cite{huppenkothen2020entrofy}, which is specifically developed to help with selecting a diverse cohort from a set of candidates. 

While the difference between format and participants' ``day job'' can help them engage in the event, it is also this difference that makes increased facilitation necessary. 
In general, how much advanced planning and organization is needed from participants is variable depending on the hackathon purpose, the communities included in the hackathon's participants, the amount of knowledge or field-specific language shared by participants. 
For example, a hackathon designed around a single purpose with one stakeholder providing the problem description and data may require less upfront effort from participants than a more open-ended design where participants can pitch their own projects.
However, such single-stakeholder approaches may also limit the creativity of the participants and may therefore reduce the value of it to some of the participants.

Designing a short, time-bounded project with a measurable outcome, forming a team and implementing teamwork structures on the extremely compressed timeline of a hackathon is often not something that all participants have much experience with. 
Before the event, facilitation typically includes guidance for the participants' preparation for the event: preparing project pitches, acquiring and cleaning any relevant data sets, identifying crucial project tasks and team member roles. 
Similarly, throughout the event, significant effort is spent on forming and facilitating project teams. 
This in particular includes mechanisms for team formation as well as for participants to change teams and for that change to happen gracefully. 
For many hackathons, where networking and learning are core goals, framing team formation and team work is crucial to create a positive learning environment. 
Especially in larger groups, it is easy for participants to get ``lost''. 
In response, organizers are often present in the space, listening in on teams and help where needed, along with scheduled check-in procedures. 

\section{Envisioning the future of hackathon research and practice}
\label{Vision}
In the following sections we discuss what we perceive to be some of the most important areas of hackathon research and practice.
The six areas are: (1) Hackathons for different purposes, 
(2) Socio-technical event design, 
(3) Scaling up, 
(4) Making hackathons equitable, 
(5) Studying hackathons, and 
(6) Hackathon goals and how to reach them.
While the listed six areas represent a synthesis of our discussions at the workshop, it is not an exhaustive list. 
Instead, we introduce the areas as an invitation to start tackling the issues we have observed in hackathon research and practice, i.e. limited exchange of best practices, limited exchange of research findings and unaddressed big, interdisciplinary challenges.

\subsection{Hackathons for different purposes}
\label{sec:vision:purposes}
A good starting point for framing hackathons for different purposes is Briscoe and Mulligan's~\cite{briscoe2014digital} definition that loosely groups hackathons as tech-centric or focus-centric. 
Tech-centric hackathon events focus on developing software and hardware using a specific technology (e.g., a hackathon aiming to promote the usage of an API and a new one). 
Focus-centric hackathons involve creating software prototypes to address a specific social issue or business objective, for instance, improving city transit systems. 
Briscoe and Mulligan's classification may be further expanded into including three categories of hackathons: corporate, educational, and civic hackathons ~\cite{gama2022}.
In addition to these categories, we may also consider research-focused hackathons, see e.g.~\cite{falk202010}.
We use this categorization as point of departure for discussing the state of the art in the next subsection.

\subsubsection{State of the art}
\textbf{Corporate hackathons} aim at broadening innovation, with participants typically motivated by learning and networking. 
They have commonly been used by IT and technology companies of all sizes, which integrate these events into their research and development activities. 
These hackathons aim to generate new ideas, early prototypes, and even business plans and can be internal or external to the organisation ~\cite{pe2022corporate, flores2018can}. 
Corporate hackathons can be either internal or external to an organisation. 
\textbf{Internal hackathons} are designed to stimulate creative thinking for the organisation and generate new ideas. 
\textbf{External hackathons} are open to participants outside the organisation and are motivated by the open-innovation paradigm by introducing new resources in crafting unique solutions. 
Both internal and external hackathons can alternate between a tech-centric or focus-centric approach or combine the two.  
These mixed events represent a strategy to support ecosystem evolution by offering a software platform and hardware for third parties to develop new products or services and encouraging outsiders to become network complementors. 
Additionally, hackathons are a way to attract and build a community of experts~\cite{granados2019collaborative}, which help to foster a broader innovation ecosystem. 

\textbf{Educational hackathons} are performed in association with teaching and learning activities, either as an initiative of a teacher or as cooperation between academia and industry -- which is sponsoring the event~\cite{nandi2016, gama2018hackathons, porras2018hackathons}. 
These hackathons are often tech-centric and can bring focus-centric approach as well:
In IT or Design courses, for instance, the hackathons become a contest for graduating students to address real-life issues in an engaging scenario that enables them to collaborate and enhance their abilities~\cite{porras2018hackathons} intensively.

\textbf{Civic hackathons} address public and societal issues organized by the public sector or non-governmental organizations. 
These hackathons focus on more socially-oriented innovation~\cite{briscoe2014digital,disalvo2014building}. 
These events are typically focus-centric hackathons, although government institutions also have been using such events to generate value from open data and APIs (a more tech-centric perspective), which different players explore (e.g., citizens, different types of companies, universities, etc.). 
These contests generally leverage the idea of government as a platform~\cite{safarov2017}.

\textbf{Research-focused hackathons} are hackathons used as a kind of research method.
In line with academia embracing research on hackathons (see section \textit{Studying hackathons} below), researchers have also organized hackathons for various research purposes, see e.g.~\cite{pe2019understanding,ghouila2018hackathons,groen2015science,falk202010}.
To mention a few examples, these purposes could be for producing and studying specific hackathon outcomes, increasing collaboration between stakeholders, and evaluating prototypes~\cite{falk202010}. 
Common for many publications using hackathons as research methods is the emphasis on how hackathons enhance and accelerate scientific collaboration, see e.g.~\cite{ghouila2018hackathons}.

\subsubsection{Research proposal}
While hackathons are evolving events, classifications can play a valuable role in analyzing these events. 
However,  understanding the different motivations and the "why" for event organizers and participants remains an open question as current evaluations and research are primarily tied to the specific type of events and their particular goals. 
Only few studies combine the different types and aims~\cite{flus2021design,briscoe2014digital}, and primarily from a design perspective. 
Generally, organizers from the different categories of hackathons, highlighted above, often mention \textit{collaboration} as a main benefit of hackathons.
Exploring this aspect from a macro perspective across different types of hackathons and focusing on motivations, expectations, and stakeholders can help us understand and organize better events. 
Furthermore, it will provide insights into how collaboration unfolds across research, industry, and education.  

\subsection{Socio-technical event design}
Technology is a key component of hackathons, which are often tech-focused as they revolve around \textit{creating} technology~\cite{briscoe2014digital}. 
Of equal importance are the technologies that organizers, mentors, participants and other stakeholders use to organize and participate in an event. 
Which technologies to use for which purpose(s) during an event is a key decision that organizers and participants need to take~\cite{nolte2020organize}. 
The importance of this decision is exacerbated by the rise of online and hybrid formats for which technology forms the foundation of organization and participation, and interactions might be facilitated entirely through technological means for virtual participants~\cite{mendes2022socio}. 
Online and hybrid events require reliable platforms for simultaneous and asynchronous interaction such as video calls and text-based chat features, virtual versions of whiteboards and note-taking facilities, and virtual spaces designed for unstructured interactions such as coffee breaks and networking events. 

Even in in-person events, technology often plays a vital role: it is used to e.g.~publicize events, support registration, foster the development and sharing of ideas ahead of and during the hackathon, support team formation, serve as a means to communicate and / or collaborate on artifacts, and submit projects. 
There are different platforms such as Devpost\footnote{\href{https://devpost.com/}{https://devpost.com/}}, Hackbox\footnote{\href{https://formidable.com/work/hackbox/}{https://formidable.com/work/hackbox/}} or Eventornado\footnote{\href{https://eventornado.com/}{https://eventornado.com/}} that offer some of those functionalities. 
While different event formats and sizes afford the use of different technologies it is clear that choices regarding technology use will affect the experience of everyone involved in organizing and participating in an event.

Online hackathons have especially seen a steep rise during the global COVID-19 pandemic~\cite{bertello2022open,temiz2020open}. 
They pose unique challenges but also provide new opportunities such as limiting the carbon footprint of travel~\cite{burtscher2020} and providing the possibility for individuals who cannot travel due to visa, funding or other issues. 
It can thus be expected that even as in-person events become more prevalent again, online events or online components of in-person events will remain, and with them, the requirement for reliable, accessible technological solutions that facilitate hackathon organization and participation.

\subsubsection{State of the art}
There is a large variety of technologies that have been proposed for use to prepare for and run hackathons like the ones mentioned before. 
Organizers often utilize a variety of technologies for different purposes. 
There are tools that are commonly used to facilitate live (Zoom, Teams, etc.)~\cite{braune2021interdisciplinary} and asynchronous interaction (Slack, Discord, etc.)~\cite{bertello2022open,fowler2020jamming} as well as sharing artifacts (Github, GDrive, etc.)~\cite{mcintosh2021hackathon,mahmoud2022one}. 
In addition, organizers often utilize websites to share the agenda or structure of an event. 
Most of our current knowledge about the use of technology in hackathons stems from studying in-person events~\cite{falk202010,taylor2018everybody,trainer2016hackathon}. 
What this work has established so far, is that the technologies that are used and the way they are used can affect the experience of participants during an event, especially in an online context. 
However, there is also a difference between technologies that organizers propose and technologies that participants end up utilizing. 
This discrepancy is driven by participants' preferences and the strength of those preferences (i.e.~technologies that a participant ``likes'' vs.~a technology that other participants ``do not mind'' using). 
Even when the same technology is being used, teams differ in the extent and the purposes for which they use a specific technology~\cite{mendes2022socio}.

Studies on the use of technology in hackathons are often limited to post-event reports of individuals that collaborated during a single or few events~\cite{bertello2022open,mendes2022socio}. 
Larger scale studies of how teams utilize technology to communicate and collaborate using e.g. trace data from communication tools are missing (one notable recent exception is~\cite{schultenparticipants}). 
It is thus unclear if the reported findings can be expected to hold across different events and event designs.
However, virtual hackathons are a relatively recent developments, thus studies on online events are just emerging~\cite{bertello2022open,mendes2022socio,happonen2021systematic} and reports on hybrid hackathons are largely non-existent at this point.

\subsubsection{Research proposal}
From the current trend towards online and hybrid hackathons, and the current state of research in this area, we propose a number of open questions that future research should address. 
In particular, it is important to know to what extent hackathon organizers can and should prescribe the technology that participants use during an event. 
Participants may be able to spot accessibility issues with platforms that organizers do not, but increased flexibility may also lead to additional confusion and fatigue in virtual participation. 
Secondly, it is currently not clear which technologies are particularly suited for the affordances of the hackathon format. 
Most technological tools are designed for commercial applications with a narrower focus than the wide variety of modes and purposes of communication commonly occurring at hackathons, and there may be no single solution that can provide all the necessary functionality. 
Finally, it is an open question how technology can be used to better support organizers of online and hybrid events, where operating the technological solutions can add significant overhead to organization and where it is currently difficult to identify technological issues and address them.

\subsection{Scaling up}
Given the time, effort and costs involved with hackathon preparations, it is only natural that the organizers want to maximize the impact by scaling it up in time (longer events) and/or size (more participants). 
The standard hackathon format is often short, usually 24 or 48 hours, and the number of participants for most events might vary between 20 and 100. 
According to Kollwitz and Dinter~\cite{kollwitz2019hack}, a short hackathon is defined as lasting less than 24h, medium as 24-72h, and long as >72h. 
The hackathon portal Devpost lists that out of 6149 hackathons on that page, 72 had more than 1000 participants, 197 had between 500 and 1000, 1787 between 100 and 500, 1290 between 50 and 500, and 2803 had less than 50 participants. 

Looking at the time dimension, there can be additional reasons for having longer hackathons. 
For example, long-term goals or larger projects require more time to be achieved and thus lend themselves to hack weeks or similar events, and for more complex or ill-defined projects time for scoping and requirements analysis needs to be factored in. 
Other hackathon projects might require asynchronous upfront work or work in between hackathon days, lending themselves to short but repeating events on the same topic. 
Some hackathons bring together communities with very different customs, language and background knowledge, where significant time must be devoted during the hackathon to community building. 
The team formation process in hackathons adds significant overhead in terms of building a shared language among the participants, and in terms of organizing the team structure. 
From an equity and inclusion perspective, it can be beneficial to organize a hackathon over multiple working days rather than a continuous 24- or 48-hour event, as participation outside of business hours and long, sustained participation can be tricky for people e.g. with caring responsibilities. 
In certain contexts, such as hackathons addressing social innovation, a wider time span may be needed, since methodologies from social sciences need more time to gain a better understanding of social contexts~\cite{ferrario2014software}.

\subsubsection{State of the Art}
When a hackathon is scaled up in the size dimension, more people usually means more teams, rather than bigger teams. This can lead to less collaboration and exchange between teams, if the organizers do not put explicit effort into facilitating this exchange. Overall facilitation of the event becomes more difficult and time-consuming for organizers. However, some communities have experimented successfully with distributed events over both a small range of time zones, as well as hackathons comprising regional pods embedded in a global organization~\cite{cameron2016brainhack}. Especially for virtual hackathons, this might provide a fruitful avenue to addressing issues around time zones.  During social distancing times in the pandemic, many hackathons aiming to crowdsource the generation of solutions to problems around COVID-19 took advantage of the online format, involving hundreds (and sometimes thousands) of participants in the same virtual event~\cite{vermicelli2020can,braune2021interdisciplinary}.

When scaling up in duration, organizers have to consider whether they want to run a single event spanning multiple, consecutive days, or rather a series of smaller events over a longer period of time. 
A single event of consecutive days might lend itself to hackathons designed for community building, sparking innovation and networking (e.g.~\cite{huppenkothen2018hack}). Examples include hackathons as a team building activity, a student event during term breaks, or to significantly advance a software project. Conversely, problems that require sustained effort within the same teams over months or years might find a regular schedule of short hack days beneficial, taking advantage of the opportunity for asynchronous work in between to prepare these hack days and drive their projects forward. Creativity research shows~\cite{boden2004creative} that this kind of distribution also gives the participants time to ``mull over'' their ideas and might lead to more creative outcomes.

As an example, in Basden et al.~\cite{basden2021performancews}, the organizers describe a hackathon which was run over seven months, with one hack day per month and asynchronous work done between the hack days. At each hack day, a new performance analysis tool was introduced that the participants, which came in existing teams, applied to their research codes. The time between hackdays was used by the teams to continue their work, asynchronously discuss problems with the tool experts, as well as for setup tasks for the next session. The feedback for this format was positive, and some teams achieved significant insights and speedups for their codes. 

Recent years have shown the advantages and disadvantages of online~\cite{mendes2022socio} and hybrid hackathons. It is still an open question whether and in which situations they provide a benefit for scaling up. While it is relatively easy to accommodate more people in an online environment in terms of logistics, as space considerations do not play a significant role, more care has to be taken to facilitate interactions between teams, and more mentors, organizers and instructors might be needed. Hackathon organizers need to be careful not to underestimate this shift in effort for scaling up online.

\subsubsection{Research proposal}
There is, to our knowledge, no rigorous research done yet on the effects and limitations of scaling up in terms of time and in terms of number of people, and how this differs of offline vs online events. 
We therefore suggest significant research effort be dedicated to examining the tradeoffs inherent in these choices.
As discussed by Falk and colleagues, shortening or lengthening the hackathon events may configure the ways in which people participate in them~\cite{falk2022shortening}.
Longer or shorter duration of hackathons may change how \textit{pace} during designing and prototyping is perceived by participants, and thereby ``influence which strategies participants pursue''~\cite{falk2022shortening} citing~\cite{grzymala2011time}.
As argued in the section above, we may ask how scaling up hackathons by organizing e.g. a series of regular schedule of short hack days may influence how participants perceive the pace of designing and prototyping?

Another interesting question to look at is scaling down: Are hackathons with teams of two persons beneficial?
Which aspects of hackathons can be retained in small-scale hackathons? 
Which effects, benefits and challenges may very short hackathons of one or two hours entail e.g. in terms of participation, accessibility, creativity and outcome?
For example, shorter schedules have been proposed as a way to to ensure broader participation of older adults~\cite{kopec2018older}.

\subsection{Making Hackathons Equitable}
As events that thrive on social interactions, hackathons tend to reproduce the power structures and discrimination of the society they are embedded in, unless carefully facilitated. 
Hackathons are perceived as typically non-inclusive events~\cite{kos2018collegiate}, and are, for example, frequented little by female participants ~\cite{decker2015understanding} where they are often subjected to different forms of discrimination~\cite{paganini2020}. 
Hackathons that explicitly welcome transgender, non-binary and gender non-conforming people are rare and represent a very small fraction of these events~\cite{kumar2019}. 
A lack of inclusion at these events means that underrepresented groups (e.g. women, LGBTQIA+) might receive fewer opportunities (e.g. learning, skill development, networking, jobs) characteristic of hackathons~\cite{warner2017}. 
A hackathon therefore cannot be considered a successful event unless it ensures equitable participation for all participants. 

Hackathons have also often been criticized for their tendency towards technological solutionism, and thus any hackathon where the outcomes will affect human lives--and especially those of traditionally minoritized groups--must ensure that stakeholders can engage with the planning and the hackathon itself~\cite{d2016towards}. 
However, representation is only the first step, and responsibility rests with the organizers to design a hackathon where participants from historically underrepresented groups are welcome.

\subsubsection{State of the art}
Equitable participation includes both logistic aspects (e.g. wheelchair access, gender-neutral bathrooms, food that respects dietary restrictions, quiet rooms) as well as aspects of facilitation (e.g. a code of conduct, facilitation structures that mitigate power dynamics in teamwork). 

Before the event, gender-neutral communication and advertisement to specific audiences are important to attract underrepresented groups~\cite{ferraz2019case}, while during the event nuances such as allowing participants to specify their preferred pronouns in identification badges can improve the sense of belonging~\cite{prado2021trans}. 
A core goal is to instill a sense in participants that they are welcome, that their experiences and skills are valued, and that they belong at the hackathon.

The logistic aspects of hackathon participation often directly depend on the chosen venue, and thus organizers should carefully evaluate whether a venue provides equitable access. 
This includes building-related features, such as wheelchair accessibility or --- in the case of age-inclusive hackathons for older participants --- simple room layouts on the first ground floor~\cite{kopec2018older}.
It also includes much broader questions around for example visa restrictions on traveling to the country where the hackathon is held, local laws, and disparities in travel funding among institutions and countries for participants. 
For a series of workshops like some hack weeks, shifting the country and venue where the workshop is located has facilitated a broader access, in combination with dedicated fundraising activities to enable travel for those participants who would otherwise not be able to attend. 
Recently, the emergence of fully virtual hackathons has provided opportunities to address logistic issues of equitable access around the space the hackathon is held in and travel funding, but has raised others (e.g. access to electricity and the necessary technology for participation, censoring of key technologies in some countries, time zones). 

As a safeguarding measure, introducing Codes of Conduct can be one way of making hackathons more equitable.
They should clearly state what constitutes unacceptable types of conduct at the event and clearly delineating (and carrying out) enforcement procedures have emerged as a key element of setting a baseline for enabling equitable participation free of discrimination and harassment~\cite{prado2021trans}. 
However, Codes of Conduct (e.g., hack code of conduct~\footnote{\href{https://hackcodeofconduct.org/}{https://hackcodeofconduct.org/}}) can only provide a baseline, and, as is the case in other team work environments, by necessity leave a large grey area of behaviour that does not strictly rise to the level of a violation, but will nevertheless make the event unwelcoming for some participants (e.g. repeatedly talking over a team member, in-group jokes, extreme competitiveness, exclusion of team members based on disparities in technical skills). 

Proactive strategies for mitigation then hinge on organizers to set the tone of the event and lead by example. 
In an analysis of 16 hackathons described in research literature, Falk et al.~\cite{falk2021hackathons} identified nine hackathons which were specifically tailored towards broadening participation.
As they discuss ``By modifying the processes and desired end goal of hackathons, researchers [and hackathon organizers] have the opportunity to include those who have been historically marginalized when considering the design of technology mediated futures''~\cite{falk2021hackathons}. 
Related to this, it has been discussed how hackathons~\cite{falk2022shortening} can be a way of facilitating participation in design processes with low investment in order to include vulnerable people ~\cite{kanstrup2018participatory}.

Hackathons often involves a sexist competitive environment not very welcoming to underrepresented groups~\cite{warner2017}.  
Fostering a competitive or collaborative ambiance is a design choice made by organizers~\cite{pe2019designing}. 
The traditional competitive hackathon format is common with incentives such as awards being offered. 
A cooperative event can be achieved when social elements are introduced -- e.g., stimulating participants to pitch project ideas or to wander around the premises and discuss with other teams -- thus helping participants from different teams to collaborate and network~\cite{pe2022corporate}. 
This supports how students, who belong to traditionally marginalized groups in computer science, tend to participate in hackathons, by embracing collaboration and non-competitive goals~\cite{kos2019understanding}.  
For the inclusion of older hackathon participants, different collaboration strategies should also be considered, such as consulting or validation~\cite{kopec2018older}.
Prado et al.~\cite{prado2021trans} interviewed transgender and gender non-conforming hackathon participants to draw a set of recommendations for more trans-inclusive hackathons: start with a gender-inclusive organizing team; foster inclusive communication; make safety visible; and showcase trans people in the event.
Other recommendations for making events more equitable are: focus more on collaboration and less on competition; stimulate development of technical and transferable skills; promote healthier habits; define an inclusive code of conduct; and include women in the organization team~\cite{paganini2021promoting}.

Successful facilitation also rests on team mentors and leads to be ready to facilitate their teams in a way that respects every team member (see also the vision of a feminist hackathon in~\cite{disalvo2014building}). 
This can involve the explicit embrace of failure as part of the hackathon, articulating the recognition that many hacks fail, and how participant may not feel competent enough.
In some events, this has also included the co-creation of a value statement for the hackathon with participants to generate buy-in.
In addition to tailoring hackathons towards broadening participation, there are also examples on hackathons in which the topic of equity is deeply embedded into the theme of the hackathon itself, inviting participants to develop ideas and prototypes that engages with the concept of \textit{Safe Spaces}\footnote{\href{https://www.unwomenuk.org/safespacesnow-hackathon}{https://www.unwomenuk.org/safespacesnow-hackathon}}; i.e. a place where people can feel confident that they will not be exposed to discrimination, criticism, harassment, or any other emotional or physical harm.

No hackathon will be perfectly equitable. 
It is thus crucially important that organizers 1) find out and recognize what went wrong at any given event, and 2) learn from those mistakes for future events (for an example, see also \href{https://github.com/MarionBWeinzierl/RS-EDI/blob/main/HackathonEDI.md}{this checklist}). 
This includes critical questions about recruitment: who was invited and solicited? Of those, who participated? Why did those who did not attend decline attendance? How did those who attend experience the hackathon event? 
Post-attendance surveys that include demographic questions can capture how experiences might have differed for participants with different backgrounds. 

\subsubsection{Research Proposal}
Despite a number of works on the topic of equity in the context of hackathons, this space remains underexplored with respect to many axes of diversity and equity, such as ethnicity, culture, socioeconomic background, gender, sexuality and neurodiversity.
As Falk et al.~\cite{falk2021hackathons} state: 
``[P]articipants’ contributions towards methodological democracy in the face of epistemological hegemony is currently underexplored and requires further engagement in future HCI [Human Computer Interaction] research utilizing modified hackathons.''
In other words, what are for instance ways in which participants have tailored and facilitated their own and others' participation according to their specific situation?

The rapid shift towards virtual and hybrid meetings in response to the COVID-19 pandemic opens up both opportunities and challenges with respect to equitable participation in hackathons that should be the focus of future work. 
In this context, one might ask whether virtual hackathons where participants are anonymous facilitate or hinder equitable participation. 
In the study of equity at hackathons, a closer collaboration with related fields in sociology and psychology exploring equitable participant selection and team work, diversity in competitive environments and anonymity in other virtual spaces may provide valuable insights and starting points for future research. 

\subsection{Studying hackathons}
In order to improve how hackathons are generally organized and run, we need to study the hackathon formats and participation. 
Studying hackathons may be valuable for not only researchers but also for organizers, who seek to evaluate and re-iterate the way they organized a hackathon.

In this section, we address the aspect of siloed research and practice by discussing how we may improve what we observe as a poor methodological fit, i.e. how we may start developing and conducting the methodology for studying hackathons by considering relevant and prior theory and thereby move towards a more mature state of theory.
A diverse range of multiple methods may help not only generating new knowledge but also advance prior theory and thereby mitigate the risk of repeating known insights related to hackathons, as a consequence of repeating what we observe as very similar studies with little changes which cannot be compared or built on.

\subsubsection{State of the art}
When diving into the research that has been done on hackathons so far it becomes evident that most works in this area focus on studying the experience of participants using qualitative methods such as post-event interviews and surveys.
Few studies also report on in-depth observations of teams during an event or focus on studying the projects that teams work on. 
Those studies focus on various aspects of the participant experience including their satisfaction with the event and the individuals involved in organizing it, their project and their team.

For qualitative research, \textit{program theory} has been discussed as a valuable concept for studying the relation between hackathon format and outcome~\cite{falk202010}, one example in this context is the work presented in~\cite{falk2021hackathons}
There are also works which utilize quantitative methods to e.g. investigated the usage of software repositories~\cite{mcintosh2021hackathon,nolte2020happens,mahmoud2022one}. 
Other researchers have also included social network analysis~\cite{gabrilove20193144}. 
Methods such as sensor based analytics and tools like smart badges which could foster the investigation of hackathons at a large scale and which have been successfully used in an educational context have not been utilized to study hackathons extensively yet~\cite{ouhaichi2021mbox}, with only very few exceptions like~\cite{lederman2015hacking}. 

We interpret this large variety of aspects that are being studied as pointing towards a lack of an agreed understanding about what is worth studying and what can be studied in the research community. 
Related to this, there is also a lack of “standardization” regarding the instruments that are used to study hackathons, which can further complicate e.g. replicability of studies.  
Most studies utilize instruments that are specifically developed to study one event or one aspect about an event.

In our experience, organizers also rarely focus on "studying" their event when planning it, with a few exceptions such as in~\cite{pe2019hackathons}. 
Instead they often mainly focus on event logistics and funding.
Often they will do what they have seen before, e.g. through attending a hackathon themselves, or what others suggest who have done it before. 
Some hackathons are also "inherited", i.e. passed on from organizer to organizer, so the "senior" will pass down their knowledge to the "junior". 
Having said that, there is also usually some development happening, i.e., small changes are made if something did not work, an exciting new tool just came up, or someone has seen something somewhere which worked well~\cite{powell2021organizing}.

\subsubsection{Research proposal}
Moving towards mature theory generally benefits ``from a mix of quantitative and qualitative data''~\cite{edmondson2007methodological}.
Based on discussions from the Lorentz Center workshop, we provide an overview of several possible qualitative and quantitative methods which readers can consider in table \ref{tab:methods}. 
The overview should be seen as a developing repertoire of methods which we may use to study hackathons, and we invite readers to partake in discussing and developing methods for how we may study hackathons.

\begin{table}[t]
\resizebox{\textwidth}{!}{%
\begin{tabular}{@{}llllllll@{}}
\toprule
\multicolumn{1}{c}{\textbf{Methods category}} &
  \multicolumn{1}{c}{\textbf{Methods}} &
  \multicolumn{3}{c}{\textbf{Hackathon stage}} &
  \multicolumn{2}{c}{\textbf{Perspective studied}} &
  \multicolumn{1}{c}{\textbf{References}} \\ \midrule
\multicolumn{1}{r}{\textit{Qualitative}} &
  \multicolumn{1}{l|}{} &
  \multicolumn{1}{c}{\textit{Before}} &
  \multicolumn{1}{c}{\textit{During}} &
  \multicolumn{1}{c|}{\textit{After}} &
  \multicolumn{1}{c}{\textit{Participants}} &
  \multicolumn{1}{c|}{\textit{Organizers}} &
   \\
 &
  \multicolumn{1}{l|}{Interviews (Contextual, structured, semi-structured etc.)} &
  x &
  x &
  \multicolumn{1}{l|}{x} &
  x &
  \multicolumn{1}{l|}{x} &
  \cite{kvale2012doing,beyer1999contextual} \\
 &
  \multicolumn{1}{l|}{Observations} &
   &
  x &
  \multicolumn{1}{l|}{} &
  x &
  \multicolumn{1}{l|}{x} &
  \cite{blomberg2017ethnographic} \\
 &
  \multicolumn{1}{l|}{Analysis (e.g. thematic)} &
   &
   &
  \multicolumn{1}{l|}{x} &
  x &
  \multicolumn{1}{l|}{x} &
  \cite{braun2012thematic} \\
\multicolumn{1}{r}{\textit{Quantitative}} &
  \multicolumn{1}{l|}{} &
   &
   &
  \multicolumn{1}{l|}{} &
   &
  \multicolumn{1}{l|}{} &
   \\
 &
  \multicolumn{1}{l|}{Output (Code, product, prototype etc)} &
   &
   &
  \multicolumn{1}{l|}{x} &
  x &
  \multicolumn{1}{l|}{} &
  \cite{nolte2020happens,imam2021secret} \\
 &
  \multicolumn{1}{l|}{RFID badges (sensor based technologies)} &
   &
  x &
  \multicolumn{1}{l|}{} &
  x &
  \multicolumn{1}{l|}{x} &
  \cite{pentland2008honest, lederman2017open} \\
 &
  \multicolumn{1}{l|}{Surveys (incl. demographics of returning participants)} &
  x &
  x &
  \multicolumn{1}{l|}{x} &
  x &
  \multicolumn{1}{l|}{x} &
  \cite{rea2014designing} \\
 &
  \multicolumn{1}{l|}{Audio/video recordings of discussions} &
   &
  x &
  \multicolumn{1}{l|}{x} &
  x &
  \multicolumn{1}{l|}{x} &
  \cite{olesen2018four} \\
 &
  \multicolumn{1}{l|}{Experience sampling method} &
   &
  x &
  \multicolumn{1}{l|}{x} &
  x &
  \multicolumn{1}{l|}{x} &
  \cite{larson2014experience} \\
 &
  \multicolumn{1}{l|}{Social media content and online chats} &
  x &
  x &
  \multicolumn{1}{l|}{x} &
  x &
  \multicolumn{1}{l|}{x} &
  \cite{bontcheva2016extracting} \\ \bottomrule
\end{tabular}%
}
\vspace{1em}
\caption{A synthesis of proposed methods for studying hackathons, compiled by the workshop participants. The provided references are a mix of literature that describe a method in a general context and literature that have applied the related method in the context of hackathons.}
\label{tab:methods}
\end{table}

Topics which could serve as the focus for future studies of hackathons are for example: How to support participants' creativity --- individually or collaboratively --- through e.g. tools, materials, or physical venue surroundings; the role of temporality on participation, design decision-making, design thinking including how different kinds of bias may be introduced, enhanced or mitigated.
Furthermore, studies of hackathons often focus on the perspective of the participants; we argue that the perspective of organizers is just as important and is often overlooked. 

As a start to avoid replication of mistakes, data from hackathon research could be shared using, for example, the FAIR (Findability, Accessibility, Interoperability, and Reusability) Data Principles, which can be used by researchers to ``enhance the reusability of their data holdings.''~\cite{wilkinson2016fair}.

In addition to these suggested methods, topics and data sharing for studying hackathons, we call for the development of shared survey instruments, used by as many organizers and participants as possible.
Generally, the aim with such shared survey instruments is to ``ensure that observed differences are in fact real differences and not an artefact of differences in the way the data were collected''~\cite{collins2003pretesting}.
Additionally, developing such instruments require interdisciplinary collaboration, in order to explore topics which is ``universally'' interesting for all fields (e.g. making hackathons equitable) but may be expressed in very different ways in different fields and contexts. 

As a point of departure for organizers for such an instrument, we suggest the following: 
For hackathon organizers who wish to study hackathons with the purpose of improving their practice, we may be inspired by a methodology described by Frick and Reigeluth as \textit{formative research}, i.e. a research methodology intended to improve theory for ``designing instructional practices or processes''~\cite{frick1999formative}.
How do organizers decide to organize their hackathons and how do they tailor the format to support participation of specific groups of maybe non-technical or even vulnerable people? 
The major concern for evaluating a certain practice is \textit{preferability} --- e.g. how a hackathon practice is better than other practices --- and consists of at least three dimensions which may be valued differently depending on the situation~\cite{frick1999formative}: 
1) \textit{Effectiveness}: How well did the practice attain the goal in the given situation?
2) \textit{Efficiency}: How effective is the practice in relation to the cost (e.g. time, money, energy)? 
3) \textit{Appeal}: How enjoyable is the practice for everyone involved?

\subsection{Hackathon goals and how to reach them}
Organizing a hackathon takes a lot of time and resources in particular on the part of the organizers. 
They thus commonly organize an event with the aim to reach specific goals as discussed before. 
Some goals can directly be achieved at an event while others might require preparation and / or follow-up activity. 
For the purpose of discussing this difference we will utilize the differentiation proposed by Falk et al.~\cite{falk2021hackathons}, using program theory as described by Hansen and colleagues ~\cite{hansen2019participatory}. 
Program theory describe how goals can relate to \textit{immediate outputs} such as artifacts that are created during hackathon events~\cite{pe2019understanding}, \textit{short or mid-term outcomes} such as a startup that is created after an event~\cite{cobham2017appfest} or \textit{long term impacts} such as establishing or growing a community~\cite{disalvo2014building}. 
Often, hackathons focus on multiple goals at the same time such as fostering the development of innovative technology that can then later be turned into products~\cite{leemet2021utilizing,nolte2018you} or fostering the integration of individuals into a community while teaching them related skills~\cite{nolte2020support,cameron2016brainhack}.

We should also note that it is not only organizers that put time and resources into preparing for and running an event. 
Other individuals such as participants, mentors, jurors, support staff or external stakeholders such as sponsors and supporters equally put time and effort into a hackathon. 
These individuals however might have very different goals than the organizers of an event. 
Their goals might then be compatible or aligned to the goals of other individuals that are involved in a hackathon or they might not~\cite{medina2019does}. 
Individuals might also have goals that are contradicting to goals of others who are involved in the same event. 
It might thus very well be that some individuals might reach their goals while others might not, even if they are involved in the same hackathon.

\subsubsection{State of the art}
Most current research and practice on hackathons focuses on \textit{immediate outputs}. 
These can be varied and include allowing participation~\cite{taylor2018everybody}, raising awareness about specific issues~\cite{hope2019hackathons}, sharing information or teaching specific subjects and practices~\cite{huppenkothen2018hack,cameron2016brainhack} or creating an artifact such as a piece of (innovative) technology~\cite{nolte2018you}. 
Works in this regard mainly report on how the design of an event can afford and affect immediate outputs. 
This includes the time that is allocated for a hackathon and the space it takes place in~\cite{trainer2016hackathon}, who is invited~\cite{cobham2017appfest}, how the design processes are supported~\cite{gama2022developers,olesen2018four} and others. 
Most work related to immediate outputs, however, relies on the perception of participants~\cite{nolte2020support} or on the perception of researchers that observing an event~\cite{pe2019understanding}. 
Whether and what participants actually learned during an event or the quality of the artifacts that they created during an event has not been extensively studied.

There are also works that discuss \textit{short or mid-term outcomes}. 
These mainly focus on hackathon projects~\cite{mcintosh2021hackathon,nolte2020happens}. 
Existing works report that few projects get continued at all~\cite{nolte2020happens,carruthers2014open} and that continuation is often left to individual participants who might or might not be in the position to carry a project forward~\cite{leemet2021utilizing,nolte2018you}. 
Prior work also reports on the difference between continuation intentions and continuation activity. 
Despite positive continuation intentions, projects are often abandoned~\cite{carruthers2014open}, only continued in the short term~\cite{nolte2020happens} or handed over to others after an event~\cite{pe2022corporate,nolte2018you}. 
There are few works that focus on short or mid-term outcomes other than projects. 
These include individual career gains~\cite{pe2022corporate,nolte2018you} and learning~\cite{porras2018hackathons}.

Few works also discuss \textit{long term impacts}. 
Existing works in this area mainly focus on hackathon projects as well especially in relation to open source. 
They report that long term project continuation is predicated by aspects such as skill diversity within a team and the intention of team members to expand the reach of their project~\cite{nolte2020happens}. 
Moreover, there are studies on the reuse of code that was created during an event. 
Findings from these studies indicate that about one third of such code gets reused in other open source projects and that the number of hackathon team members increase reuse probability~\cite{mahmoud2022one}. 
Other long term impacts such as community integration and addressing larger issues such as environmental and public health issues have not been extensively studied.

Finally there are few studies focusing on the alignment of goals between organizers and participants. 
They report that organizers and participants might not share the same goals and provide indication that the hackathon format itself might afford reaching certain goals such as networking and learning~\cite{medina2019does}.
The goals of other individuals involved in a hackathon such as mentors, jurors, support staff or external stakeholders such as sponsors and supporters have not been studied extensively so far.

\subsubsection{Research proposal}
Regarding existing work related to fostering specific goals through hackathons there are a few areas that future research could address. 
First there is a necessity to expand studies beyond immediate outputs and focus on short or mid-term as well as long term impacts. 
Moreover future work could focus on developing instruments that help assess actual impact rather than relying on the perception of individuals.

Another area that requires attention is studying the goals of individuals that are involved in preparing for and running hackathons. 
This relates to goal alignment as well as to include goals beyond those of organizers and participants.

Finally there is a lack of studies that consider hackathons in their context. 
Events are often studied out of context and organized as one-off events which are only marginally connected to other activities that e.g. a community or corporation undertakes. 
To advance the format and unlock its full potential it is necessary to consider hackathons in their larger context both when organizing and studying them.

\section{Next steps}
\label{NextSteps}
In order to facilitate more interdisciplinary sharing of best practices and collaboration on hackathon research is first of all to create \textit{awareness} that both research and practice is siloed.
Creating awareness is the purpose of this paper and we do this first and foremost by targeting a broad audience.
How can we then start sharing best practices and establish interdisciplinary research collaborations and thereby start address grander challenges which requires interdisciplinary approaches?
A first step towards this, was the Lorentz workshop that initiated this very paper, where several researchers and organizers from multiple disciplines started the conversation.
With this paper, we both call for  more interdisciplinary collaboration to address the three challenges identified in the \textit{Introduction} section and we envision that the paper can serve as a point of departure for these collaborations.
Especially the sections \textit{Hackathons in context} and \textit{Envisioning the future of hackathon research and practice} should serve as a solid foundation for guiding the maturing of future hackathon research and practice.
Our next concrete step towards breaking down the silos, is to establish and nurture an interdisciplinary community around hackathons by organizing workshops --- open to both researchers and organizers --- around hackathon research and practice.

\begin{acks}
We thank The Lorentz Center in Leiden, The Alfred P. Sloan Foundation, and the participants of the Hack the Hackathon: Shaping the Future of Hackathon Research and Practice workshop:
Abasi-Amefon Affia, 
Megan Bedell,
Amy Cannon,
Daniel Foreman-Mackey, 
Morgan Fouesneau, 
Remi Gau, 
Daniela Gawehns, 
Kevin Gott, 
Joseph Gum, 
Linda Hayden, 
Henriette Jensenius, 
Amruta Jaodand,
Meris Longmeier, 
Je’aime Powell,
Emilio Mayorga,
François Michonneau,
Audris Mockus, 
Lavinia Paganini,
Adrian Price-Whelan, 
Karla Peña Ramírez, 
Brigitta Sipőcz, 
Sarah Stone,
Myrthe Vel Tromp and Liz Vu.
\end{acks}

\bibliographystyle{ACM-Reference-Format}
\bibliography{references}

\end{document}